\documentclass[epj]{webofc}\usepackage[varg]{txfonts}\usepackage{bm}
\usepackage{booktabs}\usepackage{dcolumn}\usepackage{amsmath}
\usepackage{color}
\woctitle{QCD@Work 2016} % HEPHY-PUB 971/16
\newcolumntype{d}[1]{D{.}{.}{#1}}

\begin{document}\title{Heavy-Meson Decay Constants:\\QCD Sum-Rule
Glance at Isospin Breaking}\author{Wolfgang Lucha\inst{1}\fnsep
\thanks{\email{Wolfgang.Lucha@oeaw.ac.at}}\and Dmitri Melikhov
\inst{1,2,3}\fnsep\thanks{\email{dmitri_melikhov@gmx.de}}\and
Silvano Simula\inst{4}\fnsep\thanks{\email{simula@roma3.infn.it}}}
\institute{Institute for High Energy Physics, Austrian Academy of
Sciences, Nikolsdorfergasse 18, A-1050 Vienna, Austria\and
D.~V.~Skobeltsyn Institute of Nuclear Physics, M.~V.~Lomonosov
Moscow State University, 119991, Moscow, Russia\and Faculty of
Physics, University of Vienna, Boltzmanngasse 5, A-1090 Vienna,
Austria\and INFN, Sezione di Roma Tre, Via della Vasca Navale 84,
I-00146 Roma, Italy}\abstract{QCD sum rules for decay constants of
heavy mesons with $u$ or $d$ quark yield for $B$ mesons much less
isospin breaking than lattice QCD but good agreement for
$D$~mesons.}\maketitle

\section{Hadronic properties scrutinized from the point of view of
QCD sum rules}\label{Sec:SR}QCD sum rules \cite{SVZ} are analytic
relations between hadron features and QCD parameters emerging~upon
evaluation of correlation functions of appropriate interpolating
operators at both the hadronic level and the QCD level by
insertion of complete sets of hadron states, exploitation of
Wilson's operator product expansion, Borel transformation, and the
quark--hadron duality assumption above effective~thresholds.

In order to both raise the \emph{accuracy\/} of QCD sum-rule
predictions and estimate the \emph{systematic errors\/}
\cite{LMSAU}, some time ago we developed \cite{LMSET} a modified
formalism which takes into account the dependence~of the effective
thresholds on the Borel parameters. This enabled the
\emph{improved extraction\/} of heavy-meson decay constants $f$
\cite{LMSDC}, notably of the inequalities
$f_{B^\ast_{(s)}}<f_{B_{(s)}}$ \cite{LMSR}, later on confirmed by
lattice QCD~\cite{HPQCD}.

QCD sum rules provide in \emph{analytic\/} form the
\emph{dependences\/} of the properties of the hadrons currently
under study on the basic parameters of QCD. This allows us to
study the impact of the isospin breaking provoked by the slight
difference in the masses of the light $u$ and $d$ quarks on the
decay constants of the heavy--light mesons, by working out the
explicit relation of the decay constants to the light quark~mass.

\section{Heavy-meson decay constants: extraction from QCD sum-rule
approach}\label{Sec:C}Hence, let us try to get hold of the
dependence of our QCD sum-rule approach on the light-quark~mass.
We analyse pseudoscalar ($P_q$) and vector ($V_q$) mesons
(generically labelled $M_q\equiv P_q,V_q$) of mass $M_{M_q},$
composed of a heavy quark $Q=b,c$ and a light quark $q=u,d$ with
masses $m_Q$ and $m_q,$~respectively. In terms of suitable
interpolating heavy--light quark-current operators, that is, the
axial-vector (A)~current $A_\mu(x)\equiv\bar
q(x)\,\gamma_\mu\,\gamma_5\,Q(x)$ and the vector (V) current
$V_\mu(x)\equiv\bar q(x)\,\gamma_\mu\,Q(x),$ the \emph{decay
constants\/} $f_{M_q}$ of the mesons [with momentum $p$ and, in
the case of vector mesons, polarization vector
$\varepsilon_\mu(p)$] are defined~by\pagebreak
$$\langle0|\,A_\mu(0)\,|P_q(p)\rangle={\rm i}\,f_{P_q}\,p_\mu\
,\qquad\langle0|\,V_\mu(0)\,|V_q(p)\rangle=f_{V_q}\,M_{V_q}\,
\varepsilon_\mu(p)\ .$$The decay constants may be inferred from
the correlators of two of such currents
$J_\mu(x)=A_\mu(x),V_\mu(x)$:$$\Pi^{(J)}_{\mu\nu}(p)={\rm
i}\int{\rm d}^4x\exp({\rm i}\,p\,x)\, \langle0|\,T\!
\left(J_\mu(x)\,J_\nu^\dag(0)\right)|0\rangle\ ,\qquad J={\rm
A},{\rm V}\ .$$The QCD sum rules providing, for the mesons
$M_q=P_q,V_q,$ their masses $M_{M_q}$ and decay constants
$f_{M_q}$ receive both purely perturbative contributions,
represented by dispersion integrals of spectral densities
$\rho_J(s,m_Q,m_q,\alpha_{\rm s}),$ given by series expansions in
powers of the strong coupling $\alpha_{\rm s},$ and
non-perturbative contributions
$\widehat\Pi^{(J)}_N(\tau,m_Q,m_q),$ parametrized by vacuum
condensates; generically, they assume the form
$$f_{M_q}^2\,M_{M_q}^{2\,N}\exp\left(-M_{M_q}^2\,\tau\right)
=\int_{(m_Q+m_q)^2}^{s_{\rm eff}(\tau)}{\rm
d}s\,s^N\exp(-s\,\tau)\,\rho_J(s,m_Q,m_q,\alpha_{\rm s})
+\widehat\Pi^{(J)}_N(\tau,m_Q,m_q)\ ,\qquad J={\rm A},{\rm V}\
.$$The integer $N$ relates to the Lorentz nature of the employed
interpolating operator: the axial-vector~and vector currents
$A_\mu(x)$ and $V_\mu(x)$ imply $N=1,$ whereas, for pseudoscalar
currents, we would get $N=2.$

The innovative change introduced, in Refs.~\cite{LMSET}, to the
concept of QCD sum rules simply consists in realizing (and
accepting) that, in general, the effective threshold $s_{\rm eff}$
will depend on the Borel variable $\tau$ (the parameter of
dimension of inverse mass squared arising upon Borel
transformation): $s_{\rm eff}=s_{\rm eff}(\tau).$ Our favourite
possibility for acquiring information on the actual $\tau$
behaviour of $s_{\rm eff}(\tau)$ is to fit our QCD sum-rule
predictions of the masses of the mesons under study to their
experimentally measured values. It suffices to model $s_{\rm
eff}(\tau)$ by polynomial ans\"atze of rather low order $n,$ with
expansion coefficients~$s^{(n)}_j$:$$s^{(n)}_{\rm eff}(\tau)=
\sum_{j=0}^ns^{(n)}_j\,\tau^j\ .$$By its central value and
half-width, the spread of results for linear ($n=1$), quadratic
($n=2$), and cubic ($n=3$) $s^{(n)}_{\rm eff}(\tau)$ behaviour
enables an estimate of some decay constant and its systematic
uncertainty~\cite{LMSET}.

We should achieve our goals by just imitating the application of
our algorithm \cite{LMSET} to the analysis of heavy-meson decay
constants in related studies \cite{LMSDC,LMSR} based on
cutting-edge QCD contributions \cite{SD,JL,G+}. In order to
optimize our predictions' perturbative performance \cite{JL}, we
use quark masses defined by the modified minimal-subtraction
($\overline{\rm MS}$) renormalization scheme. Table~\ref{Tab:P}
shows all numerical ingredients.

\begin{table}[h]\centering\caption{Numerical parameter values
entering in QCD sum rules for heavy-meson decay constants
\cite{LMSDC,JL,PDG,FLAG}.}\label{Tab:P}\begin{tabular}{lcc}
\toprule\multicolumn{2}{c}{Operator-product-expansion parameter}&
Numerical input value\\\midrule$u$-quark $\overline{\rm MS}$ mass&
$m_u(2\;\mbox{GeV})$&$(2.3^{+0.7}_{-0.5})\;\mbox{MeV}$\\$d$-quark
$\overline{\rm MS}$ mass&$m_d(2\;\mbox{GeV})$&
$(4.8^{+0.5}_{-0.3})\;\mbox{MeV}$\\$s$-quark $\overline{\rm MS}$
mass&$m_s(2\;\mbox{GeV})$&$(93.8\pm2.4)\;\mbox{MeV}$\\$c$-quark
$\overline{\rm MS}$ mass&$m_c(m_c)$&$(1275\pm25)\;\mbox{MeV}$\\
$b$-quark $\overline{\rm MS}$ mass&$m_b(m_b)$&
$(4247\pm34)\;\mbox{MeV}$\\Strong coupling&$\alpha_{\rm s}(M_Z)$&
$0.1185\pm0.0006$\\Light-quark condensate&$\langle\bar
q\,q\rangle(2\;\mbox{GeV})$&$-[(267\pm17)\;\mbox{MeV}]^3$\\
Strange-quark condensate&$\langle\bar s\,s\rangle(2\;\mbox{GeV})$&
$(0.8\pm0.3)\times\langle\bar q\,q\rangle(2\;\mbox{GeV})$\\Gluon
condensate&$\displaystyle\left\langle\frac{\alpha_{\rm
s}}{\pi}\,G\,G\right\rangle$&$(0.024\pm0.012)\;\mbox{GeV}^4$\\
\bottomrule\end{tabular}\end{table}

\section{Summary of procedures, results, insights, conclusions, and
outlook \cite{LMSIB}}\label{Sec:S}Allowing the light-quark mass
$m_q$ to assume any value within the interval $0\le m_q\le m_s$
and defining~on this domain a function $f_M(m_q)$ by the outcomes
of our QCD sum rule for the decay constant of a~meson $M$ with
light quark of just this mass $m_q,$ we minimize uncertainties by
analyzing the difference quotient\begin{equation}\label{Eq:R}
\frac{f_M(m_d)-f_M(m_u)}{m_d-m_u}\ ,\qquad
\left.M\right|_{m_q=m_u,m_d,m_s}=M_{u,d,s}\ ,\end{equation}or the
\emph{slope\/} of $f_M(m_q)$ in the vicinity of
$\overline{m}\equiv(m_u+m_d)/2,$ requiring us to specify some
$m_q$ dependences:\begin{itemize}\item For the light-quark
condensate, $\langle\bar q\,q\rangle,$ we assume a linear $m_q$
dependence from its value $\langle\bar u\,u\rangle\approx
\langle\bar d\,d\rangle$ at $\overline{m}\equiv(m_u+m_d)/2
=3.5^{+0.7}_{-0.2}\;\mbox{MeV}$ \cite{PDG} up to the strange-quark
condensate, $\langle\bar s\,s\rangle,$ at $m_s=93.8\;\mbox{MeV}.$
\item For the mass $M_M(m_q)$ of any heavy--light [$\bar
Q\,q$] meson, we allow for a linear $m_q$ dependence from the
experimental \cite{PDG} nonstrange-meson mass $M_{M_{ud}}$ up to
the associated strange-meson mass $M_{M_s}$~\cite{PDG}:
$$M_M(m_q)=M_{M_{ud}}+\frac{m_q-\overline{m}}{m_s-\overline{m}}
\left(M_{M_s}-M_{M_{ud}}\right).$$\end{itemize}

\begin{figure}[b]\centering\begin{tabular}{cc}
\includegraphics[scale=0.33788,clip]{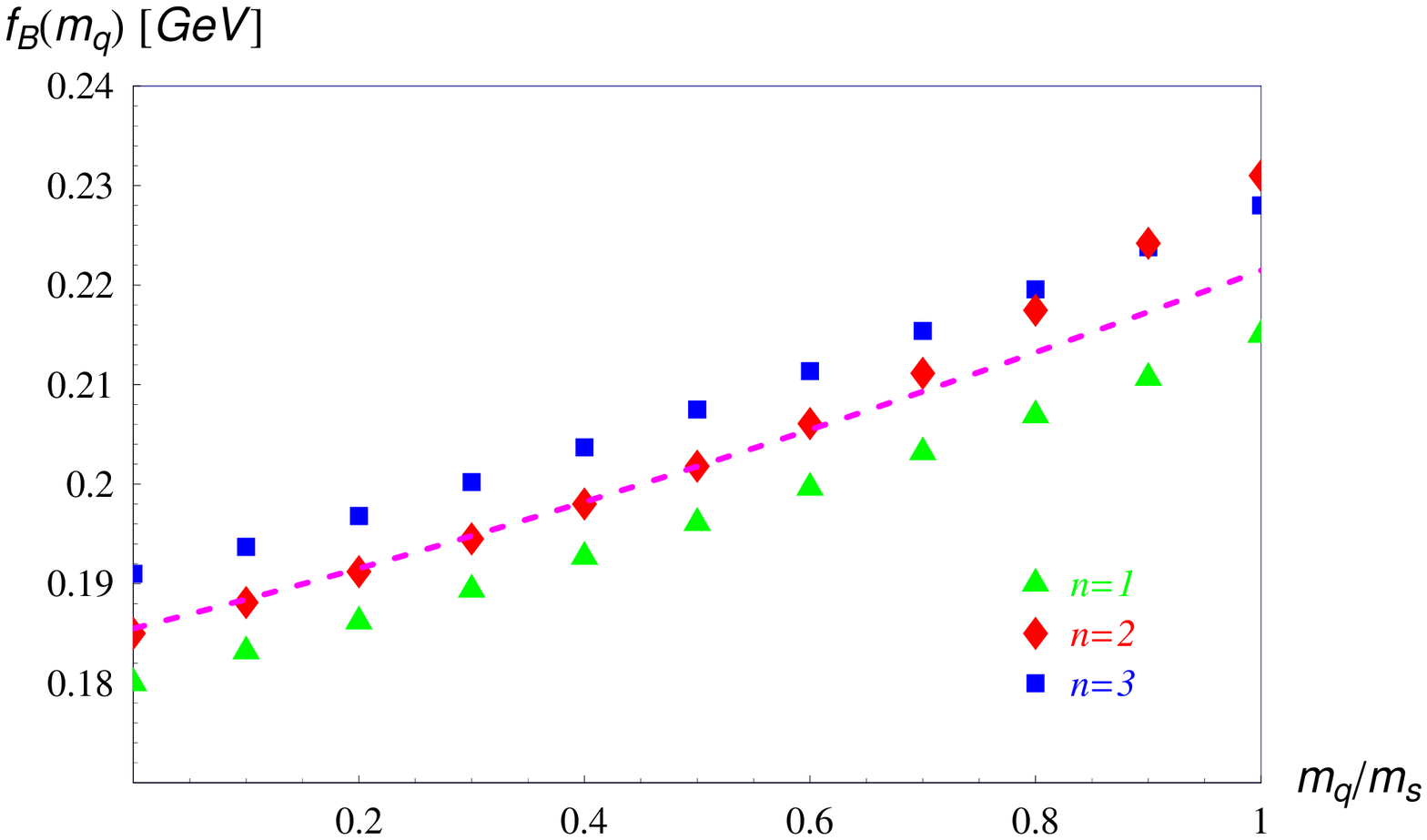}&
\includegraphics[scale=0.33788,clip]{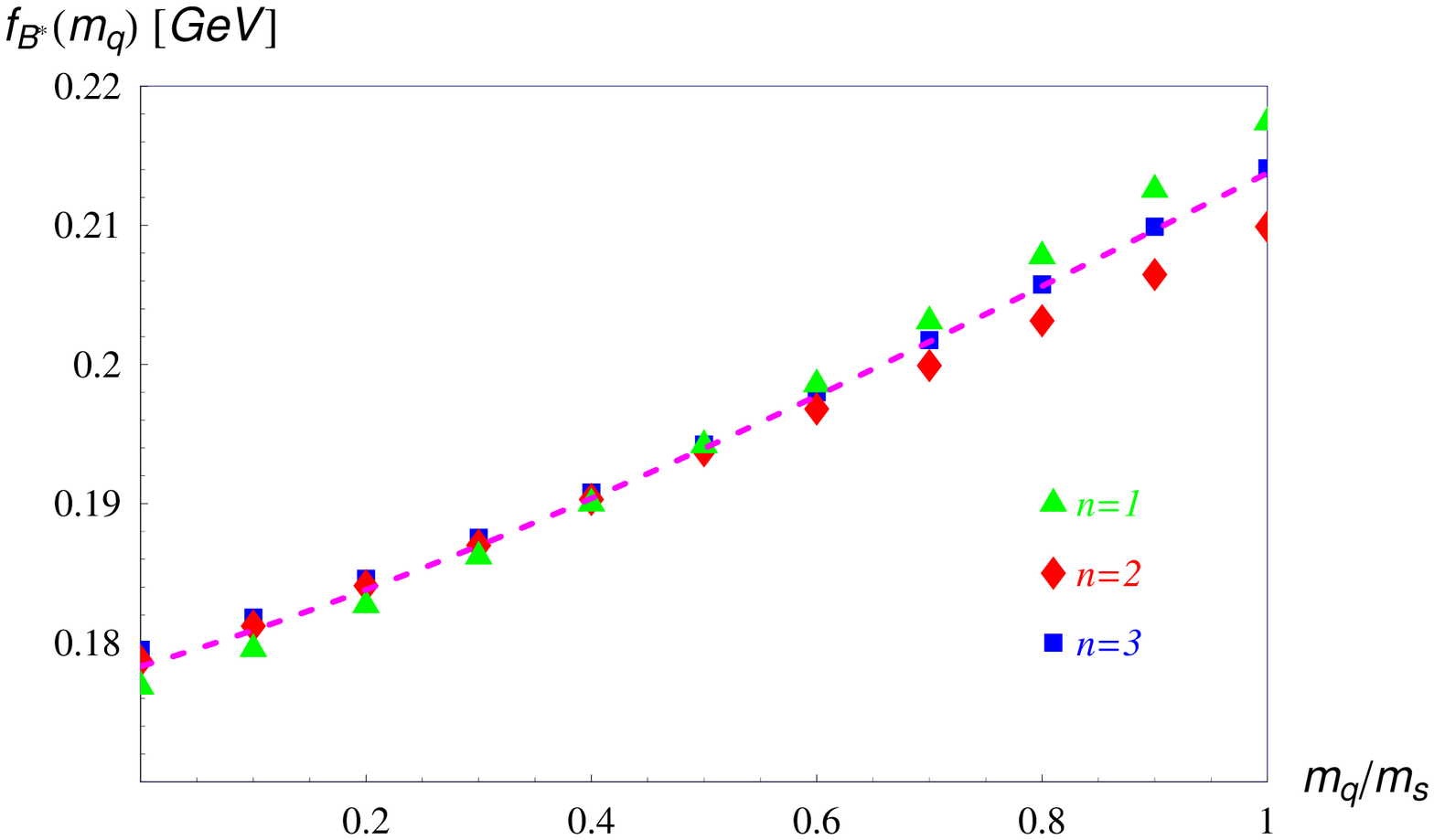}\\(a)&(b)\\[1ex]
\includegraphics[scale=0.33788,clip]{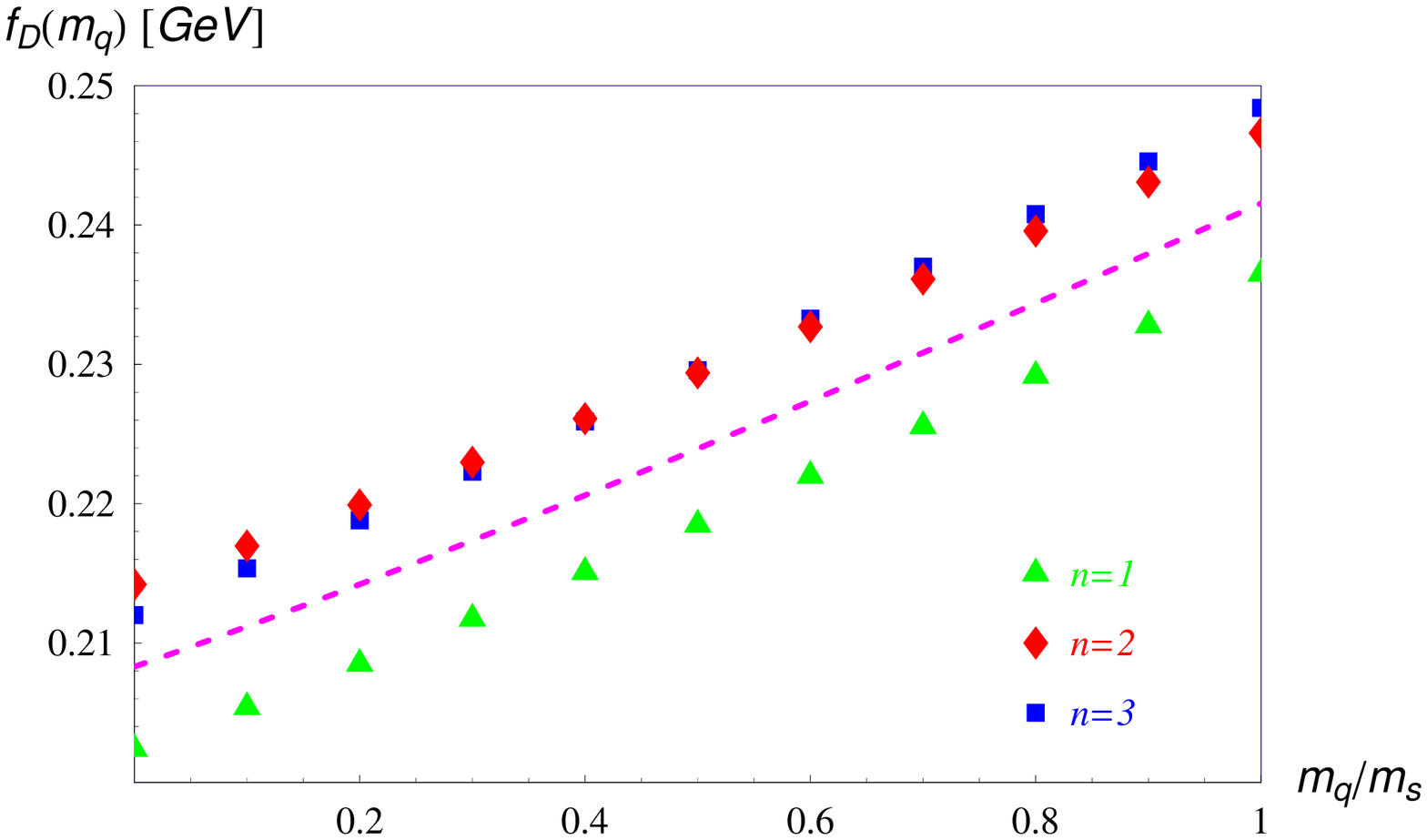}&
\includegraphics[scale=0.33788,clip]{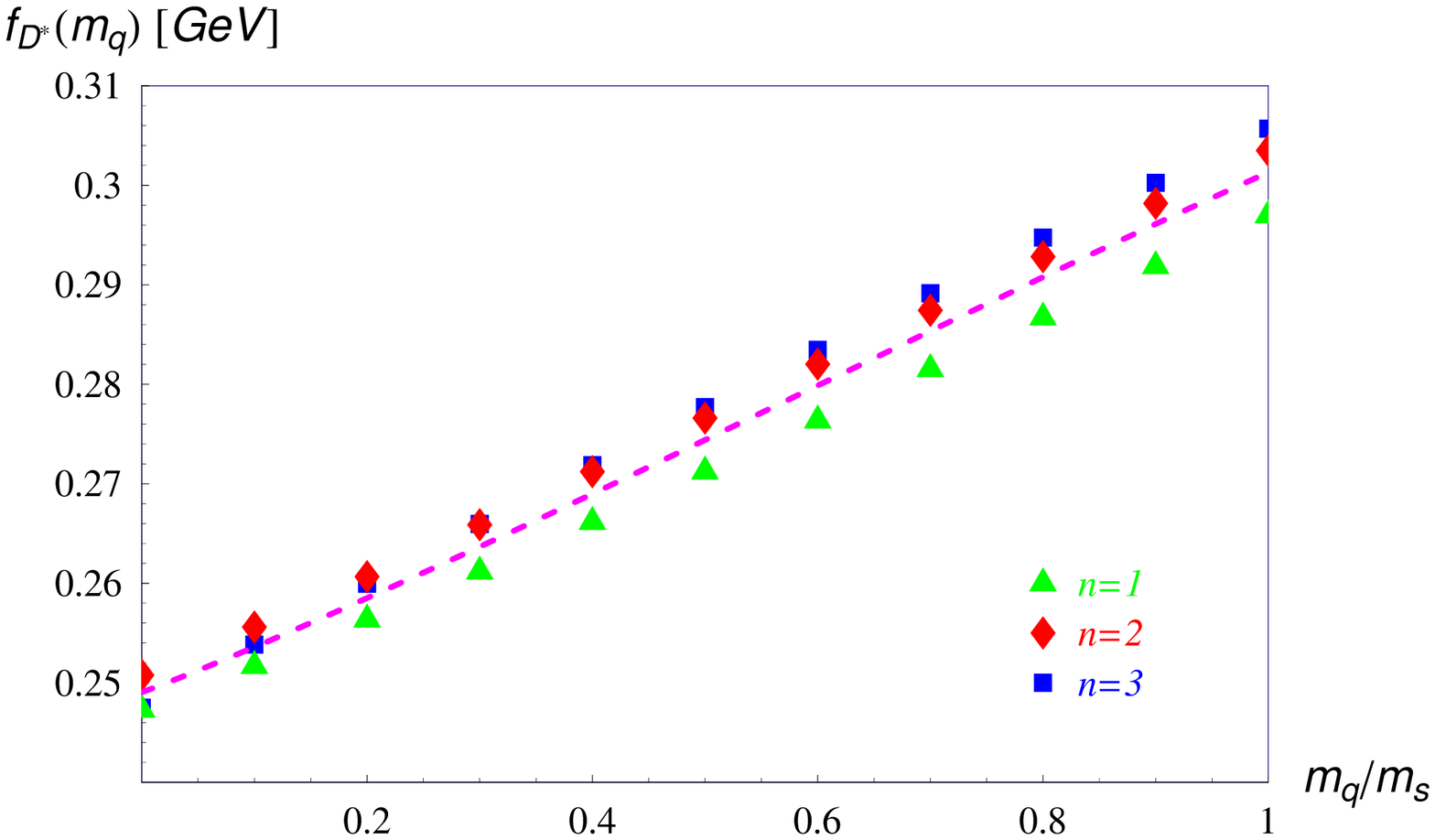}\\(c)&(d)
\end{tabular}\caption{Decay constants $f_M$ of both [$\bar b\,q$]
beauty mesons $B$ (a) and $B^\ast$ (b) and [$\bar c\,q$] charmed
mesons $D$ (c) and $D^\ast$ (d): dependence on the light quark
mass $m_q$ predicted by QCD sum rules \cite{LMSET} employing, for
the effective~threshold $s_{\rm eff}(\tau),$ polynomial ansatzes
of order $n=1$ (green triangles
\textcolor{green}{$\blacktriangle$}), $n=2$ (red diamonds
\textcolor{red}{$\blacklozenge$}), and $n=3$ (blue squares
\textcolor{blue}{$\blacksquare$}). The dashed magenta line
indicates the position of the center of the band spanned by these
individual-$n$~extractions.}\label{Fig:n}\end{figure}

\begin{figure}[t]\centering\begin{tabular}{cc}
\includegraphics[scale=0.33788,clip]{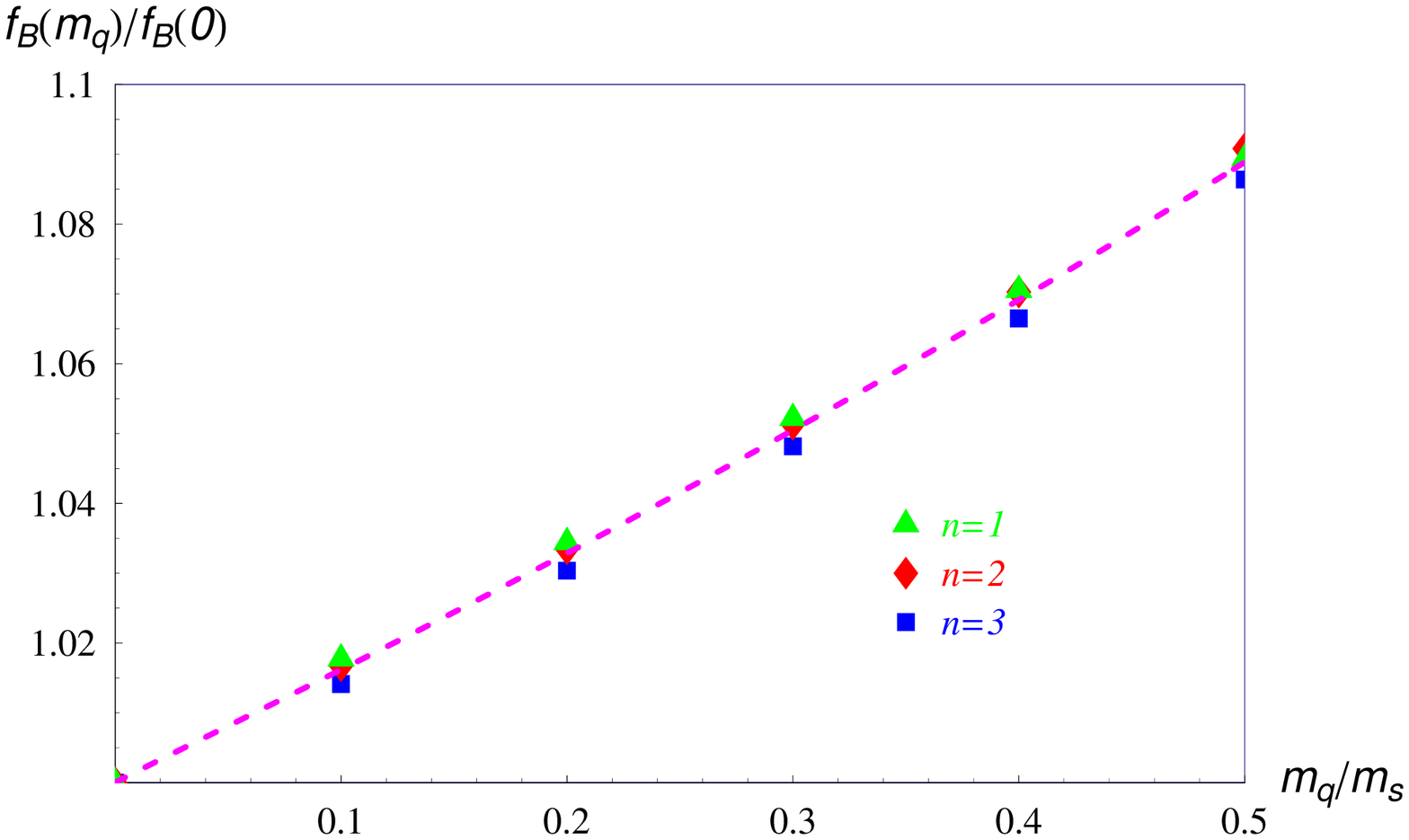}&
\includegraphics[scale=0.33788,clip]{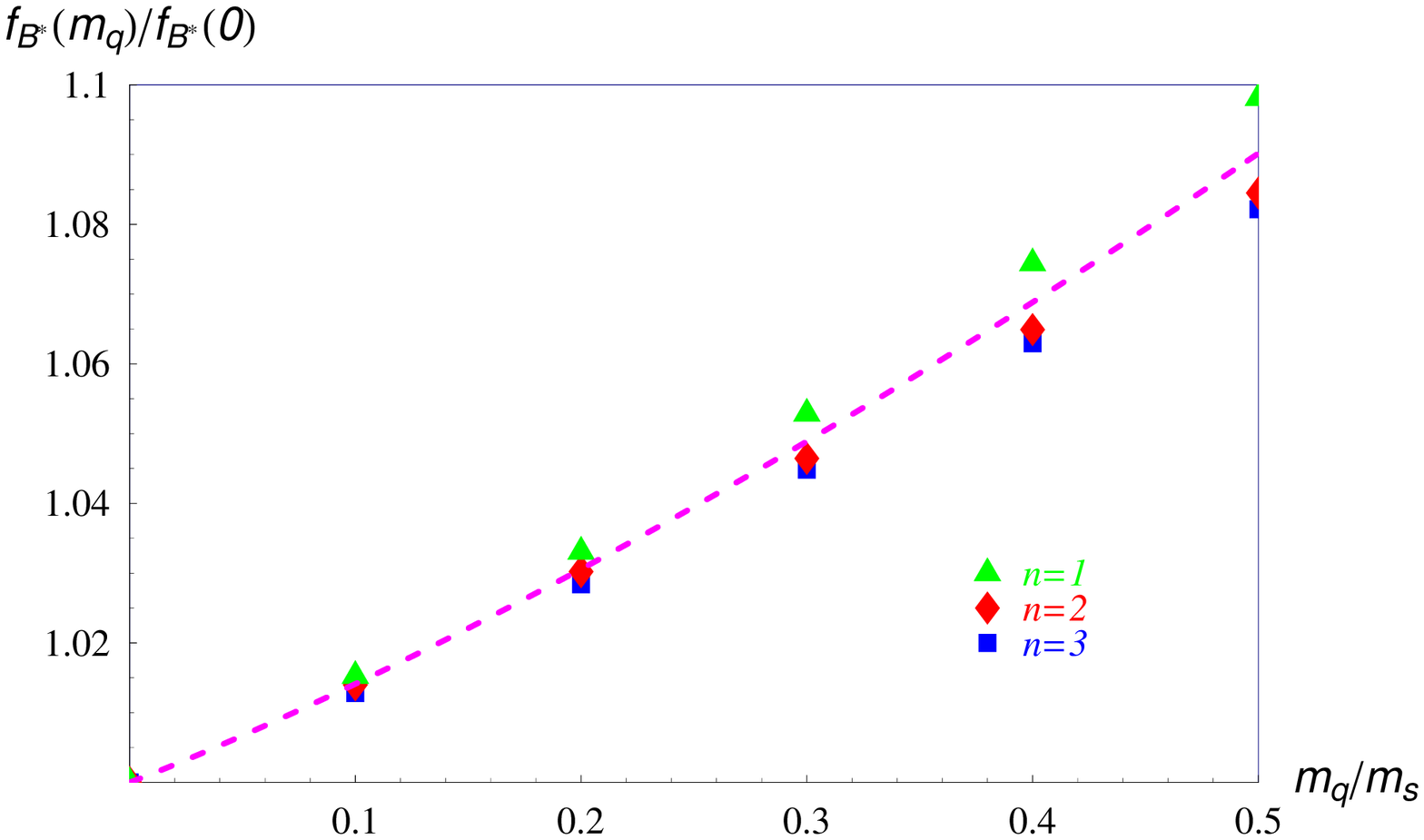}\\(a)&(b)\\[1ex]
\includegraphics[scale=0.33788,clip]{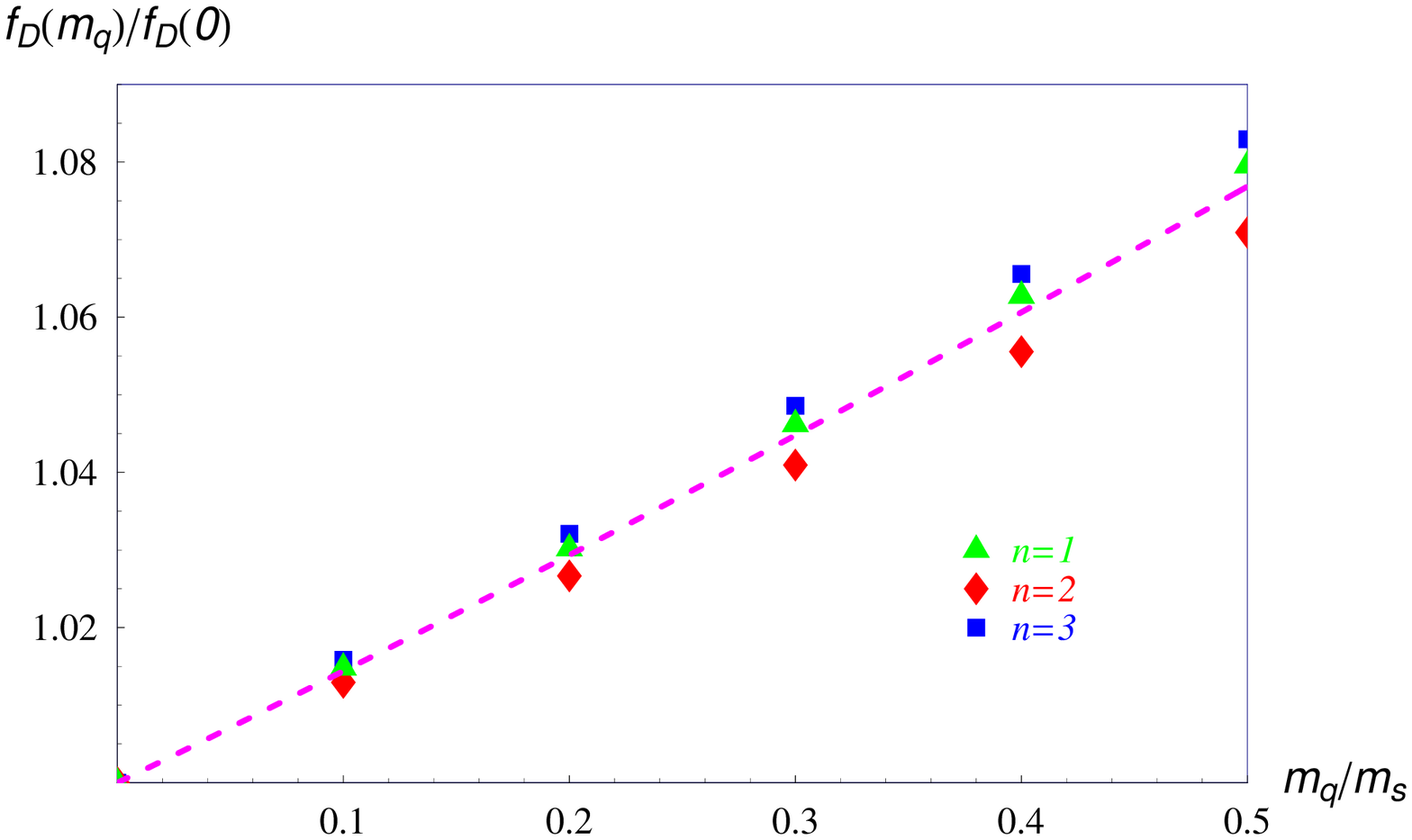}&
\includegraphics[scale=0.33788,clip]{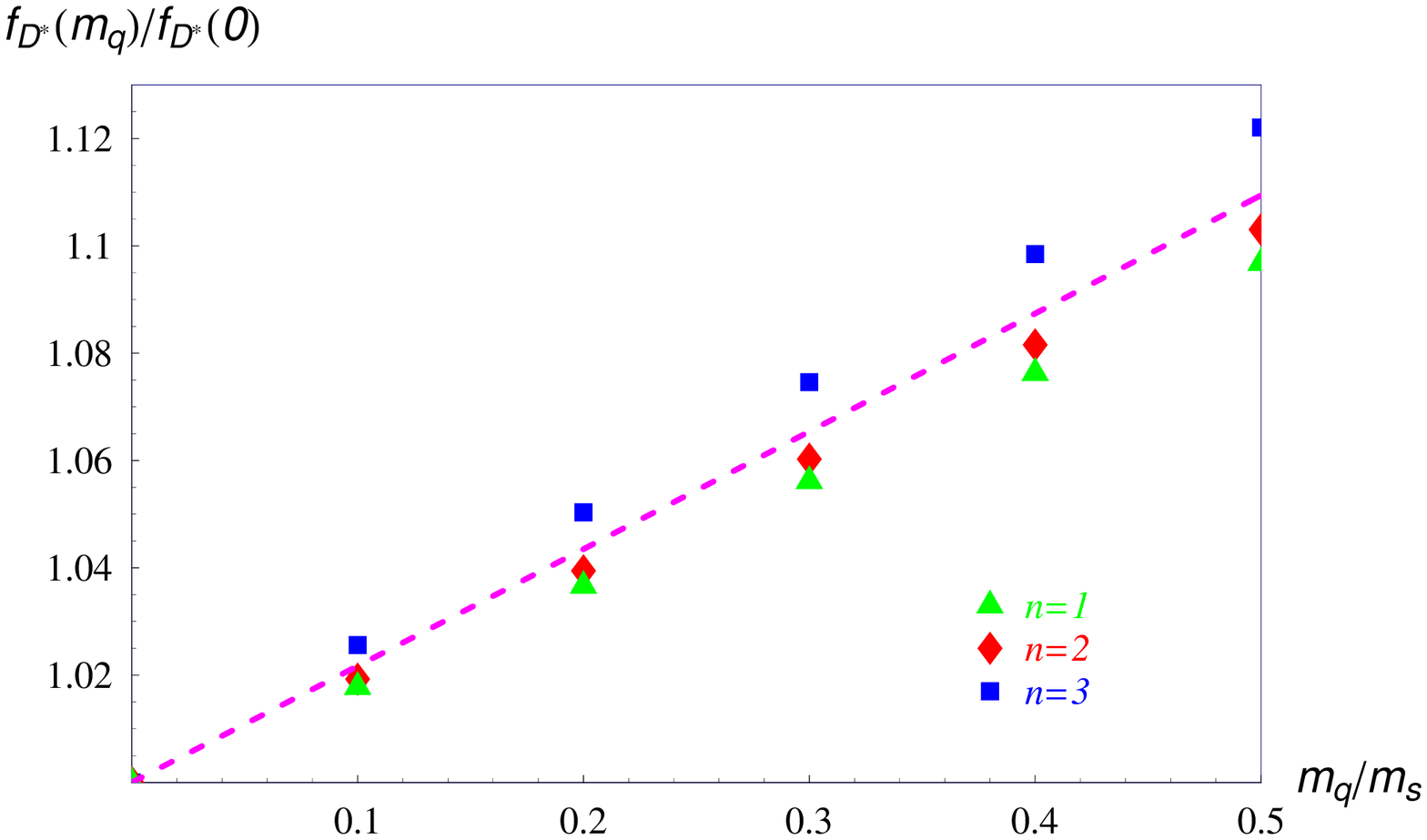}\\(c)&(d)
\end{tabular}\caption{Decay-constant ratios,
$f_M(m_q)/f_M(0),$ of the bottom mesons $B$ (a) and $B^\ast$ (b)
and the charmed mesons $D$ (c) and $D^\ast$ (d): dependence on the
light-quark mass $m_q$ from QCD sum rules \cite{LMSET} relying,
for the effective~threshold $s_{\rm eff}(\tau),$ on polynomial
ans\"atze of order $n=1$ (green triangles
\textcolor{green}{$\blacktriangle$}), $n=2$ (red diamonds
\textcolor{red}{$\blacklozenge$}) or $n=3$ (blue squares
\textcolor{blue}{$\blacksquare$}). The dashed magenta line
indicates the position of the center of the band spanned by these
individual-$n$~extractions.}\label{Fig:r}\end{figure}

\begin{table}[b]\centering\caption{Difference of the decay
constants $f_{M_q}$ for $d$- vs.~$u$-quark mesons signalling
isospin breaking, normalized to $f_M(0),$ for
$M=B^{(\ast)},D^{(\ast)},$ and numerical values of the
coefficients $r_{\ell,1}$ in the parametrization (\ref{Eq:RP}) of
these ratios.}\label{Tab:M}
\begin{tabular}{ld{1.10}d{1.10}d{1.13}}\toprule Meson $M_q$&
\multicolumn{1}{c}{$r_\ell$}&\multicolumn{1}{c}{$r_1$}&
\multicolumn{1}{c}{$\displaystyle\frac{f_{M_d}-f_{M_u}}{f_M(0)}$}\\
\midrule$B$&0.011\pm0.008&0.181\pm0.003&(4.1\pm0.4)\times10^{-3}\\
$B^*$&0.024\pm0.003&0.19\pm0.02&(3.6\pm0.3)\times10^{-3}\\[1ex]
$D$&0.006\pm0.015&0.156\pm0.010&(3.8\pm0.4)\times10^{-3}\\
$D^*$&0.002\pm0.010&0.22\pm0.010&(5.7\pm1.2)\times10^{-3}\\
\bottomrule\end{tabular}\end{table}

Figure~\ref{Fig:n} shows the decay-constant function $f_M(m_q)$
for all $M=B^{(\ast)},D^{(\ast)},$ whereas Fig.~\ref{Fig:r}
illustrates \pagebreak nicely the reduction of the systematic
errors, when normalizing $f_M(m_q)$ to their values $f_M(0)$ at
$m_q=0.$ Albeit, as physical quantities, the decay constants
$f_{M_q}$ must not depend on the renormalization scale $\mu,$
inevitable truncations of perturbative expansions induce
artificial $\mu$ dependences of any QCD sum-rule results, raising
the errors. We find $f_{M_q}$ from the ranges \cite{LMSDC,LMSR}
$1<\mu\;(\mbox{GeV})<3$ for charmed~mesons and
$3<\mu\;(\mbox{GeV})<5$ for beauty mesons. For each of the mesons
analyzed, $f_M(m_q)$ may be parametrized~as
\begin{equation}\frac{f_M(m_q)}{f_M(0)}=1+r_\ell\,\frac{m_q}{m_s}
\log\frac{m_q}{m_s}+r_1\,\frac{m_q}{m_s}+\cdots\ ,\qquad
M=B^{(\ast)},D^{(\ast)}\ .\label{Eq:RP}\end{equation}Table
\ref{Tab:M} reveals, for each meson $M_q,$ the values of the
coefficients $r_{\ell,1}$ and the decay-constant difference
$f_{M_d}-f_{M_u},$ which satisfies
$(f_{M_d}-f_{M_u})/(m_d-m_u)>0,$ tantamount to $f_{M_d}>f_{M_u}$
for all $M=B^{(\ast)},D^{(\ast)}$ \cite{LMSIB}. Upon taking into
account all the uncertainties induced by the QCD parameters and
the systematic ones inherent to the QCD sum-rule technique, we
arrive at the following $f_{B^{(\ast)},D^{(\ast)}}$ decay-constant
differences:\begin{align*}
f_{B^0}-f_{B^\pm}&=(0.79\pm0.14)\;\mbox{MeV}\ ,&
f_{B^{*0}}-f_{B^{*\pm}}&=(0.65\pm0.10)\;\mbox{MeV}\ ,\\
f_{D^\pm}-f_{D^0}&=(0.78\pm0.13)\;\mbox{MeV}\ ,&
f_{D^{*\pm}}-f_{D^{*0}}&=(1.41\pm0.42)\;\mbox{MeV}\ .\end{align*}
Comparing with related lattice-QCD outcomes \cite{LIB}, our
predictions for the decay-constant differences are, for the $B$
mesons, lower than their lattice-QCD counterparts but agree, in
the case of the $D$ mesons, with lattice-QCD results. Our findings
are of similar size as those got by lattice QCD for the $K$
mesons. In Ref.~\cite{LMSIB}, we present a more detailed analysis,
including the uncertainties of the light-quark masses.

\vspace{2.49ex}\noindent{\bf Acknowledgement.} D.~M.\ was
supported by the Austrian Science Fund (FWF): Project P29028-N27.

\end{document}